\documentstyle[epsfig]{aipproc2}
\begin{document}

\pagestyle{plain}

\title{Nuclear Forces and Nuclear Structure
\thanks{Invited Talk presented at the {\it Nuclear Structure '98}
conference, Gatlinburg, Tennessee, August 10-15, 1998.}
}

\author{R. Machleidt}

\address{Department of Physics, University of Idaho,
 Moscow, Idaho 83844\\
Electronic address: machleid@uidaho.edu}

\maketitle

\begin{abstract}
After a historical review, I present
the progress in the field of realistic NN
potentials that we have seen in recent years.
A new generation of very quantitative
(high-quality/high-precision) NN potentials
has emerged.
These potentials will serve as reliable input for
microscopic nuclear structure calculations and will allow for
a systematic investigation of off-shell effects.
The issue of three-nucleon forces is also discussed.
\end{abstract}

\section*{Introduction}

The goal of nuclear physics is to explain the properties of atomic
nuclei in fundamental terms---where the stress is on {\it fundamental}.
The debate then begins with what the fundamental ingredients
(or `first principles') are from which we should start.
Nuclear physics is strong interaction physics, so, based 
upon the Standard Model, it should be quarks exchanging gluons.
However, in the spirit of the currently fashionable effective
field theories, one may argue that any theory is effective,
and what theory is appropriate
depends on the energy scale. 
In fact, Weinberg~\cite{Wei79} pointed out
20 years ago that an effective field theory of nucleons and
mesons that observes the same symmetries as QCD is equivalent
to QCD.  This allows us to consider mesons and nucleons 
as the basic items, on our energy scale. 
The  spin-off from the mesons is the nuclear force.

Thus, it is suggested that the appropriate fundamental 
ingredients that nuclear structure should be based upon are nucleons and
the `fundamental' nucleon-nucleon (NN) interaction (created by meson exchange).
The above discussion also provides clear guideline for 
how to extend the model if it fails. 
One may introduce explicit meson degrees of freedom and
one may complement the nucleon by meson-nucleon resonances 
(e.g., the $\Delta(1232)$ isobar); moreover,
one may take into account the quark substructure of nucleons.
However, it should be stressed that the starting 
point---besides being fundamental---must also be simple.
Nucleons interacting via the two-nucleon force is the simplest
of all basic pictures. Only where we have clear evidence that
this frame work is insufficient may we extend it.
For most problems of conventional nuclear structure, we
do not have such evidence at this time.

Starting from nucleons and the bare two-nucleon force,
any nuclear many-body problem one may consider is exactly
defined. However, due to limitations in computing power,
the problem can be solved exactly only up to $A=8$~\cite{Pie98},
and since the needed computing time goes up as factorial
of $A$, there is little hope that we will ever get beyond
$A=12$. Therefore, in most problems of nuclear structure
and reactions, the first step is to derive an effective
interaction from the bare NN potential. Usually, this
involves Brueckner theory or some variation/extension
thereof~\cite{KO90,HKO95}. However, it is important to stress
that---no matter what many-body theory or model 
is used---this step has to be done in a {\it parameter-free}
way. Finally, the effective interaction may be applied
in the shell model, yielding predictions for nuclear properties.

Following this scheme (which is known as {\it microscopic}
nuclear structure), the predictions depend only on two
fundamental items: the bare NN potential and the nuclear
many-body theory/model applied. Thus, the comparison of the
predictions with the data will allow for conclusions concerning
the NN potential (off-shell) and the appropriate nuclear many-body
theory. In other words, we will learn something
about the foundations of our profession, which is our ultimate
goal.

In this conference, we have many contributions that deal with
particular effective interactions, shell-model applications,
predicted and measured nuclear properties.
It is the purpose of this contribution to fill-in on one of the
fundamental ingredients that is at the outset of all this,
namely, the NN interaction. As it turns out, we have had substantial progress
in recent years and---as a result of this---the starting
conditions for {\it microscopic} nuclear structure
calculations are better than ever.

To put the recent advances into perspective,
I will start with a brief historical review and then present
the new developments. Artificially, I will draw the line
between `history' and recent progress/new developments in the year of 1990.

\section*{Historical Perspective}

Here, I will give a brief summary of the status in the field 
of nuclear forces before 1990.\footnote{A more comprehensive historical
account is given in Refs.~\cite{Mac89,ML94}.}
This will make it easier for us to appreciate the progress
of the past eight years.
Historically, one may distinguish between three decades of work
on the meson theory of nuclear forces, namely, the 60's, 70's, and
80's. Each decade is characterized by progress in a particular
sector of meson theory, namely, the one-boson-exchange model, the two-pion
exchange contribution to the NN interaction, and contributions
beyond $2\pi$, respectively.

\subsection*{The One-Boson-Exchange Model}

The first 
quantitative meson-theoretic models for the NN interaction
were the one-boson-exchange potentials (OBEP).
They emerged right after the experimental 
discovery of heavy mesons in the early 1960's.
In general, about
six non-strange bosons with masses below 1 GeV are taken into account:
the pseudoscalar mesons $\pi (138)$ and $\eta (549)$, 
the vector mesons $\rho (769)$ and $\omega (783)$, and two scalar bosons
$a_0/\delta (983)$ and $\sigma (\approx 550)$. The first particle 
in each group 
is isovector (isospin $I=1$) while the second is isoscalar ($I=0$).
The pion provides the tensor force, which is reduced at short range
by the $\rho$ meson.
The $\omega$ creates the spin-orbit force and the short-range repulsion,
and the $\sigma$ is responsible for the intermediate-range attraction.
Thus, it is easy to understand why a model which includes at least four
mesons can reproduce the major properties of 
the nuclear force.\footnote{The
interested reader will find a detailed, pedagogical 
introduction into the OBE model in
sections~3 and 4 of Ref.~\cite{Mac89}.}
 
A classic example for an OBEP is the 
Bryan-Scott potential published in 1969~\cite{BS69}. 
Since it is suggestive to think of a potential as a function of $r$
(where $r$ denotes the distance between the centers of the two interacting
nucleons), the OBEP's of the 1960's where represented as 
local $r$-space potentials.
To reduce the original one-meson-exchange Feynman amplitudes
to
such a simple form, drastic approximations have to be applied.
The usual method is to expand the amplitude in terms of $p^2/M$
and keep only terms up to first order 
(and, in many cases, not even all of them).
Commonly, this is called the 
{\it nonrelativistic OBEP}.
Besides the suggestive character of a local function of $r$,
such potentials are easy to apply in $r$-space calculations.
However, the original potential (Feynman amplitudes) is nonlocal,
and, thus, has a very different off-shell behaviour than its
local approximation. Though this does not play a great role in two-nucleon
scattering, it becomes important when the potential
is applied to the nuclear few- and many-body problem. 
In fact, it turns out that the original nonlocal potential
leads to much better predictions in nuclear structure than the
local approximation (see below).

Historically, one must understand that
after the failure of the pion theories in the 1950's,
the one-boson-exchange (OBE) model was considered a great
success in the 1960's. 
However, there are conceptual and quantitative problems with this model.

A questionable point of the OBE model is the fact that it has to 
introduce a 
scalar-isoscalar $\sigma $-boson in the mass range 
500--700 MeV, the existence of which is controversial 
(for a current status report, see Ref.~\cite{PDG96}).
Furthermore, the model is 
restricted to single exchanges of bosons that are `laddered' in an unitarizing 
equation. Thus, irreducible multi-meson exchanges, which may be quite 
sizable (see below), are neglected. 

Quantitatively, a major drawback of the {\it nonrelativistic}
 OBE model is its failure to 
describe certain partial waves correctly.
The Bryan-Scott nonrelativistic OBE
potential 
predicts
$^1P_1$ and $^3D_2$
phases substantially above the data.~\cite{foot1}

An important advance during the 1970's  has been the development of the
{\it relativistic OBEP}~\cite{ROBEP}. In this model, the full, relativistic
Feynman amplitudes for the various one-boson-exchanges
are used to define the potential. These nonlocal expressions do not
pose any numerical problems  when used in momentum space.\footnote{
In fact, in momentum space, the application of a nonlocal
potential is numerically
as easy as using the momentum-space representation of a local
potential.}
The quantitative deficiencies of the nonrelativistic OBEP disappear
when the non-simplified, relativistic, nonlocal 
OBE amplitudes are used.

The {\it Nijmegen potential}~\cite{NRS78}, published in 1978, 
 is a nonrelativistic $r$-space OBEP.
As a latecomer, it is one of the most
sophisticated examples of its kind.
It includes all non-strange mesons of the pseudoscalar, vector, and
scalar nonet.
Thus, besides the six mesons mentioned above, 
the $\eta'(958)$, $\phi(1020)$, and
$S^*(993)$ are taken into account.
The model also includes the dominant J=0 parts of the Pomeron and
tensor ($f$, $f'$, $A_2$) 
trajectories, which essentially lead to repulsive
central Gaussian potentials. 
For the Pomeron (which from
today's point of view may be considered as a multi-gluon exchange) 
a mass of 308 MeV is assumed.
However, this potential is still defined in terms
of the nonrelativistic local approximations to the OBE amplitudes.
Therefore, it shows exactly the same problems 
as its ten years older counterpart, the Bryan-Scott potential: 
the Nijmegen potential fails to predict the $^1P_1$ and $^3D_2$ phase shifts 
correctly to the same extent as 
the Bryan-Scott potential.
In addition, the Nijmegen potential overpredicts the $^3D_3$ phase shifts
by more than a factor of two~\cite{foot1}, a problem the models of the 1960's
did not have.  The inclusion of more bosons does obviously
not improve the model and is, thus, unnecessary.
It is more important to calculate the one-boson exchange amplitudes
correctly and without the local approximations.

\subsection*{The $2 \pi$-Exchange Contribution to the NN Interaction}

In the 1970's, work on the meson theory of the nuclear force focused on the 
$2\pi$-exchange contribution to the NN interaction to replace the 
$\sigma$-boson. One way to calculate these contributions is by means 
of dispersion relations. 
Around 1970, many groups throughout the 
world were involved in this approach; we mention here, in particular,
the Stony Brook~\cite{JRV75} 
and the Paris~\cite{Vin73} groups. 
These groups could show that the intermediate-range part of the nuclear force
is, indeed, decribed correctly by the $2 \pi$-exchange 
as obtained from dispersion
integrals.

To construct a complete potential, the Stony Brook as well as the Paris
group complemented their $2 \pi$-exchange contribution by one-pion-exchange
(OPE) 
and $\omega$ exchange. In addition to this, the Paris potential~\cite{Lac80}
 contains a 
phenomenological short-range potential for $r< 1.5$ fm.

In microscopic nuclear structure calculations, the off-shell
behavior of the NN potential is important. 
The fit of NN potentials to two-nucleon data 
fixes them only on-shell.
The off-shell behavior cannot, in principle, 
be extracted from two-body data.
Theory could 
determine the off-shell nature of the potential;
however, not every theory can do that. Dispersion theory relates
observables (equivalent to on-shell $T$-matrices) to observables;
e.~g., $\pi N$ data to $NN$ data.
Thus, dispersion theory cannot, in principle, 
provide any off-shell information.
The Paris potential is based upon dispersion theory; thus, the off-shell
behavior of this potential is not determined by the underlying
theory.
On the other hand, every potential does have an off-shell behavior.
When undetermined by theory, then the off-shell behavior is a
silent by-product of the parametrization chosen to fit the
on-shell $T$-matrix, with which the potential is identified.
In summary, due to its basis in dispersion theory, the 
off-shell behaviour of the Paris potential is 
not derived on theoretical grounds.
This is a serious drawback when it comes to the question of how to
interpret nuclear structure results obtained with this potential.

\subsubsection*{The Field-Theoretic Approach to $2 \pi$-Exchange}

In a field-theoretic picture,
the interaction between mesons and baryons is described by
effective Lagrangians.
The NN interaction can then be derived in terms of field-theoretic
perturbation theory.
The lowest order (that is, the second order in 
terms of meson-baryon interactions) are
the one-boson-exchange diagrams, which are easy to calculate.

More difficult (and more numerous) are the irreducible two-meson-exchange
(or fourth order) diagrams.
It is reasonable to start
with the contributions of longest range. These are the graphs that
exchange two pions. 
There are two complications that need to be taken into account:
meson-nucleon resonances and meson-meson scattering.
The lowest $\pi N$ resonance, the so-called $\Delta$
isobar with a mass of 1232 MeV, gives rise
to one set of diagrams. 
The other set involves $\pi\pi$ interactions.
Since it is well established that two pions in relative $P$-wave 
form a resonance (the $\rho$ meson), there is no problem with
$\rho$ exchange. However, the existence of
a proper resonance in $\pi\pi-S$-wave below 900 MeV is
controversial~\cite{PDG96}.
In any case, there are strong correlations between two pions in 
$S$-wave at low energies.
Durso {\it et al.}~\cite{DJV80} have shown that these correlations
can be described in terms of a broad mass distribution of about
$600\pm 260$ MeV, which in turn can be approximated by
a zero-width scalar-isoscalar ($\sigma$) boson of mass 550 MeV~\cite{MHE87}.

In summary, two approaches are available for calculating 
the $2\pi$-exchange
contribution to the NN interaction:
dispersion theory (Paris~\cite{Vin73,Lac80}) 
and field theory (Bonn~\cite{MHE87}).
One can compare the predictions by the two approaches 
with each other as well as with the data. 
For this purpose, one looks into
the peripheral partial waves of NN scattering. 
By and large, one finds satisfactory agreement~\cite{MHE87}.

\subsection*{$\pi \rho$-Contributions and the Bonn Potential}
A model consisting of $\pi+2\pi+\omega$
describes the peripheral partial waves quite satisfactory.
However, when proceeding to lower partial waves
(equivalent to shorter internucleonic distances),
this model generates too much attraction.
This is true for the dispersion-theoretic (Paris)
as well as the field-theoretic (Bonn) result.
Obviously, further measures have to be taken at short range
to arrive at a quantitative
model for the NN interaction. 
It is at this point that the philosophies of the Bonn and Paris groups diverge.

The Paris group decided to give up meson theory at this stage and to
describe everything that is still missing by phenomenology. 
Thus, they added a phenomenological short-range potential for $r<1.5$ fm
which requires many parameters (see Table~1, below) 
that do not allow for a clear-cut
physical interpretation.
This short-range potential affects 
the S-,  P-, and D-waves of NN scattering which are,
therefore, largely a product of phenomenology in the Paris model.

In contrast, the Bonn group continued to consider further irreducible
two-meson exchanges. The next set of 
diagrams to be considered are the exchanges
of $\pi$ and $\rho$. 
As it turns out, these contributions very accurately take care  of the
discrepancies that remained between theory and 
experiment~\cite{MHE87}.

Thus, the $\pi\rho$
contributions provide the short-range repulsion
which was still missing. It is important to note that the $\pi\rho$
contributions have only one free parameter, namely the cutoff for
the $\rho N \Delta$ vertex. The other parameters involved occur also
in other parts of the model and were fixed before (like the $\pi NN$
and $\rho NN$ coupling constants and cutoff parameters). 
Notice also that the $\pi N\Delta$
and $\rho N \Delta$ coupling constants are not free parameters, since
they are related to the corresponding $NN$ coupling constants by
$SU(3)$.

In summary, a proper meson-theory for the NN interaction should
include the irreducible diagrams of $\pi\rho$ exchange.
This contribution has only one free parameter and
makes comprehensive short-range phenomenology
unnecessary. Thus, meson theory can be truly tested
in the low-energy NN system.

In the 1970's and 80's,
a field-theoretic model for the NN interaction was developed at the 
University of Bonn. This model
consists of single $\pi$, $\omega$, and $a_0/\delta$ exchange, 
the field-theoretic $2\pi$
model, and $\pi\rho$ diagrams, 
as well as
a few more irreducible $3\pi$ and $4\pi$  diagrams (which are not very 
important, but indicate convergence of the diagrammatic expansion). 
This quasi-potential 
has become known as the `Bonn full model'~\cite{MHE87}.
It has 12 parameters which are the coupling constants and cutoff masses
of the meson-nucleon vertices involved.
With a reasonable choice for these parameters, a very satisfactory
description of the NN observables up to 300 MeV is achieved
(see Table~1).
Since the goal of the Bonn model was to put meson theory
to a real test, 
no attempt
was ever made to minimize the $\chi ^2$ of the fit of the NN data.
Nevertheless, the Bonn full model shows the smallest $\chi^2$
for the fit of the NN data among the traditional models.

\begin{table}[t]
\caption{Comparison of some typical meson-theoretic nucleon-nucleon
models of the pre-1990 era.}
\begin{tabular}{|l|c|c|c|}
\hline
\hline
              & Nijmegen(1978)~\cite{NRS78}  
& Paris Potential(1980)~\cite{Lac80} & Bonn full model(1987)~\cite{MHE87} \\
\hline
\hline
{\it \# of free parameters} 
                     &    15 & $\approx 60$   & 12 \\
\hline
{\it Theory includes:} &&& \\
OBE terms     &     Yes        &  Yes  &  Yes     \\
$2\pi$ exchange&    No         &  Yes  &  Yes      \\
$\pi\rho$ diagrams& No         &  No   &  Yes     \\
Relativity   &      No         &  No   &  Yes      \\
\hline
{\it $\chi^2$/datum for fit of} &&& \\
{\it world NN data:} &&& \\
$pp$ data    & 2.06             &  2.31  &  1.94   \\
$np$ data    &6.53              &  4.35  &  1.88   \\
$pp$ and $np$ data&5.12              &  3.71  &  1.90   \\
\hline
\hline
\end{tabular}
\end{table}

\subsection*{Summary}
In Table 1, we give a summary and an overview of the theoretical
input of some representative meson-theoretic NN models 
of the pre-1990 era.
Moreover, this table also lists the $\chi^2$/datum 
for the fit of the world NN data below 300 MeV laboratory energy,
which is 5.12, 3.71, and 1.90 for the Nijmegen~\cite{NRS78},
Paris~\cite{Lac80}, and Bonn~\cite{MHE87} potentials, respectively. 
This compact presentation, typical for a table, makes
it easy to grasp one important point: The more seriously and consistently
meson theory is pursued, the better the results.
This table and its trend towards the more comprehensive meson models
is the best proof for the validity of meson theory in the low-energy
nuclear regime.
To show this fact, which is of fundamental importance to our field,
was the major achievement of the pre-1990 era.

While all models considered in Table~1 describe the proton-proton ($pp$)
data well (with $\chi^2$/datum $\approx 2$), some models have a problem
with the neutron-proton ($np$) data (with $\chi^2$/datum $\approx 4-6$).
For the case of the Paris potential (and, in part, for the Nijmegen potential)
this is due to a bad 
reproduction of the $np$ total cross section data.
When the latter data are ignored, the Paris potential fits $np$ as well as $pp$.
The Nijmegen and the Paris potential predict too large $np$
total cross sections because their $^3D_2$ phase shifts are
too large~\cite{foot1}.

This finishes the review of the developments concerning
the NN interaction up to the late 1980's.
Next, we turn to the progress made in the 1990's.

\section*{Recent Progress in NN Potentials}

In the 1990's, the focus has been on the 
quantitative aspect of the NN potentials.
Even the best NN models of the past 
fit the NN data typically with a $\chi^2$/datum $\approx 2$ or more.
This is still substantially above the perfect
$\chi^2$/datum $\approx 1$. 
To put microscopic nuclear structure theory to a reliable test,
one needs a perfect NN potential such that discrepancies in the
predictions cannot be blamed on a bad fit of the NN data by
the potential.

To construct perfect NN potentials one needs, first, a perfect
NN analysis.
About a decade ago, the Nijmegen group embarked on a program
to substantially improve NN phase shift analysis.
Finally, in 1993, they could publish a phase-shift analysis
of all proton-proton and neutron-proton data below
350 MeV laboratory energy with a $\chi^2$ per datum of 0.99
for 4301 data~\cite{Sto93}.
Based upon this analysis, 
new `high-precision' or `high-quality' NN potentials
have been constructed which, similar to the analysis, fit
the NN data with a $\chi^2$/datum $\approx 1$.

I will first review the Nijmegen NN analysis and then 
the new potentials.

\subsection*{The Nijmegen NN Analysis}

In spite of the huge NN database available today, 
conventional phase shift analyses are by no means perfect.
For example, the phase shift solutions obtained
by Bugg~\cite{BB92} or the VPI group~\cite{Arn92}
typically have a $\chi^2/$datum of about
1.4, for the energy range 0--425 MeV. 
This may be due to inconsistencies in the data
as well as deficiencies in the constraints applied in the analysis.
In any case, it is a matter of fact that
within the conventional phase shifts analysis, in which the lower partial
waves are essentially unconstrained, a better fit cannot be achieved.

To further improve NN analysis, the Nijmegen group took two 
decisive measures~\cite{Sto93}.
First, they `pruned' the data base; i.e., 
they scanned very critically the world NN data base 
(all data in the energy range 0-350 MeV laboratory energy 
published in a regular physics journal
between January 1955 and December 1992) and eliminated all data that
had either an improbably high $\chi^2$ 
(more than three standard deviations off) or
an improbably low $\chi^2$; 
of the 2078
world $pp$ data below 350 MeV  
1787 survived the scan, and of the 3446 $np$ data 2514 survived.
Second, they introduced
sophisticated, semi-phenomenological model assumptions 
into the analysis. Namely,
for each of the lower partial waves ($J\leq 4$) a
different energy-dependent potential is adjusted
to constrain the energy-dependent analysis.
Phase shifts are obtained using these potentials in a Schroedinger
equation. From these phase shifts
the predictions for the observables are calculated including the $\chi^2$
for the fit of the experimental data. This $\chi^2$ is then minimized
as a function of the parameters of the partial-wave potentials.
Thus, strictly speaking, the Nijmegen analysis is a {\it potential analysis};
the final phase shifts are the ones
predicted by the `optimized' partial-wave
potentials.

In the Nijmegen analysis,
each partial-wave potential consists of a short- and a long-range
part, with the separation line at $r=1.4$ fm.
The long-range potential $V_L$ ($r>1.4$ fm) 
is made up of an electromagnetic part $V_{EM}$
and a nuclear part $V_N$: 
\begin{equation}
V_L=V_{EM}+V_N 
\end{equation}
The electromagnetic interaction
can be written as
\begin{equation}
V_{EM}(pp)=V_C+V_{VP}+V_{MM}(pp)
\end{equation}
for proton-proton scattering and
\begin{equation}
V_{EM}(np)=V_{MM}(np)
\end{equation}
for neutron-proton scattering,
where $V_C$ denotes an improved Coulomb potential
(which takes into account the lowest-order relativistic corrections
to the static Coulomb potential and includes contributions
of all two-photon exchange diagrams);
$V_{VP}$ is the vacuum polarization potential,
and $V_{MM}$ the magnetic moment interaction.

The nuclear long-range potential $V_N$
consists of the local one-pion-exchange (OPE) tail $V_{\pi}$
(the coupling constant $g_\pi$
being one of the parameters used to minimize
the $\chi^2$) multiplied by a factor $M/E$
and the tail of the heavy-boson-exchange (HBE)
contributions of the Nijmegen78 potential~\cite{NRS78} $V_{HBE}$,
enhanced by a factor of 1.8 in singlet states; i.~e.
\begin{equation}
V_N=\frac{M}{E}\times V_{\pi}(g_\pi,m_\pi) + f(S)\times V_{HBE}
\end{equation}
with $f(S=0)=1.8$ and $f(S=1)=1.0$,
where $S$ denotes the total spin of the two-nucleon system.
The energy-dependent factor $M/E$ (with $E=\sqrt{M^2+q_0^2}$, 
 $\; q_0^2=MT_{lab}/2$) takes into account
relativity in a `minimal' way, damping the nonrelativistic
OPE potential at higher energies.

As indicated, $V_{\pi}$  depends on the $\pi NN$ coupling constant
$g_\pi$
 and the pion mass $m_\pi$, which gives rise to charge dependence.
For $pp$ scattering, the OPE potential is
\begin{equation}
V_{\pi}^{pp}=V_{\pi}(g_{\pi^0},m_{\pi^0})
\end{equation}
with $m_{\pi^0}$ the mass of the neutral pion.
In $np$ scattering, we have to distinguish between $T=1$ and $T=0$:
\begin{equation}
V^{np}_{\pi}(T)=-V_{\pi}(g_{\pi^0},m_{\pi^0})
+(-1)^{T+1} 2 V_{\pi}(g_{\pi^\pm},m_{\pi^\pm})
\end{equation}

The partial-wave short-range potentials ($r \leq 1.4$ fm)
are energy-dependent square-wells (see Figs.~2 and 3 of Ref.~\cite{Sto93}).
The energy-dependence of the depth of the square-well is 
parametrized in terms of up to three parameters per
partial wave.
For the states with $J\leq 4$, there are a total of 39 such parameters
(21 for $pp$ and 18 for $np$)
plus the pion-nucleon coupling constants ($g_{\pi^0}$ and $g_{\pi^\pm}$).

In the Nijmegen $np$ analysis, the $T=1$ $np$ phase shifts are calculated
from the corresponding $pp$ phase shifts
(except in $^1S_0$ where an independent analysis is conducted) 
by applying corrections due to
electromagnetic effects and charge dependence of OPE. 
Thus, the $np$ analysis determines $^1S_0(np)$ and the
$T=0$ states, only.

In the combined $pp$ and $np$ analysis~\cite{Sto93}, the fit for 
1787 $pp$ data and 2514 $np$ data below 350 MeV
results in the `perfect' $\chi^2/$datum = 0.99.

\subsection*{The New High-Precision NN Potentials}

Based upon the Nijmegen analysis and the (pruned)
Nijmegen data base, new charge-dependent NN potentials were
constructed in the early/mid 1990's.
The groups involved and the names of their new creations are,
in chronological order:
\begin{itemize}
\item
Nijmegen group~\cite{Sto94}: Nijm-I, Nijm-II, and Reid93 potentials.
\item
Argonne group~\cite{WSS95}: $V_{18}$ potential.
\item
Bonn group~\cite{MSS96}: CD-Bonn potential.
\end{itemize}
All these potentials have in common that they use about 45 parameters and
fit the pruned Nijmegen data
base with a $\chi^2$/datum $\approx 1$ (cf.\ Table 2, below).
The larger number of parameters (as compared to previous OBE
models) is necessary to achieve the very accurate fit.

Concerning the theoretical basis of these potential, 
one could say that they are all---more or less---constructed
`in the spirit of meson theory' (e.g., all potentials include
the one-pion-exchange contribution). However, there are
considerable differences in the details leading to considerable
off-shell differences among the potentials.

To explain these details and differences in a systematic way,
let me first sketch the general scheme for the derivation
of a meson-theoretic potential.

One starts from field-theoretic 
Lagrangians for meson-nucleon coupling, which are essentially
fixed by symmetries. Typical examples for such Langrangians are:
\begin{eqnarray}
{\cal L}_{ps}&=& -g_{ps}\bar{\psi}
i\gamma^{5}\psi\varphi^{(ps)}
\\
{\cal L}_{s}&=& g_{s}\bar{\psi}\psi\varphi^{(s)}
\\
{\cal L}_{v}&=&g_{v}\bar{\psi}\gamma^{\mu}\psi\varphi^{(v)}_{\mu}
+\frac{f_{v}}{4M} \bar{\psi}\sigma^{\mu\nu}\psi(\partial_{\mu}
\varphi_{\nu}^{(v)}
-\partial_{\nu}\varphi_{\mu}^{(v)})
\end{eqnarray}
where
$ps$, $s$, and $v$ denote pseudoscalar, scalar, and vector
couplings/fields, respectively.

The lowest order contributions to the nuclear force from the above
Lagrangians are the second-order Feynman diagrams
which, in the center-of-mass frame of the two interacting nucleons, produce
the amplitude:
\begin{equation}
{\cal A}_{\alpha}(q',q) =
\frac{\bar{u}_1({\bf q'})\Gamma_1^{(\alpha)} u_1({\bf q}) P_\alpha
\bar{u}_2(-{\bf q'})\Gamma_2^{(\alpha)} u_2(-{\bf q})}
{(q'-q)^2-m_\alpha^2} \; ,
\end{equation}
where $\Gamma_i^{(\alpha)}$ ($i=1,2$) are vertices
derived from the above Lagrangians, $u_i$ are Dirac spinors representing
the nucleons, and $q$ and $q'$ are the nucleon relative momenta
in the initial and final states, respectively; $P_\alpha$
divided by the denominator is  the
meson propagator.

The simplest meson-exchange model for the nuclear force is
the one-boson-exchange (OBE) potential~\cite{Mac89} which sums over
several second-order diagrams, each representing 
the single exchange of a different boson, $\alpha$: 
\begin{equation}
V({\bf q'},{\bf q}) =
\sqrt{\frac{M}{E'}}
\sqrt{\frac{M}{E}}
\sum_\alpha i{\cal A}_{\alpha}({\bf q'},{\bf q}) 
F_{\alpha}^2({\bf q'},{\bf q}) \; .
\end{equation}
As is customary, we include form factors,
$F_{\alpha}({\bf q'},{\bf q})$, applied to the
meson-nucleon vertices, and a square-root factor
$M/
\sqrt{E'E}$
(with $E=\sqrt{M^2+{\bf q}^2}$
and $E'=\sqrt{M^2+{\bf q'}^2}$; $M$ is the nucleon mass).
The form factors regularize the amplitudes for large momenta
(short distances) and account for the extended structure of
nucleons in a phenomenological way.
The square root factors make it possible to cast the 
unitarizing,
relativistic, three-dimensional Blankenbecler-Sugar (BbS) equation
for the scattering amplitude 
[a reduced version of the four-dimensional Bethe-Salpeter (BS)
equation] 
into a form
which is identical to the (nonrelativistic) Lippmann-Schwinger
equation~\cite{Mac89}. Thus, Eq.~(11) defines a relativistic
potential which can be consistently applied in conventional, nonrelativistic
nuclear structure.

\begin{figure}[t]
\vspace*{1.5cm}
\hspace*{5cm}\epsfig{file=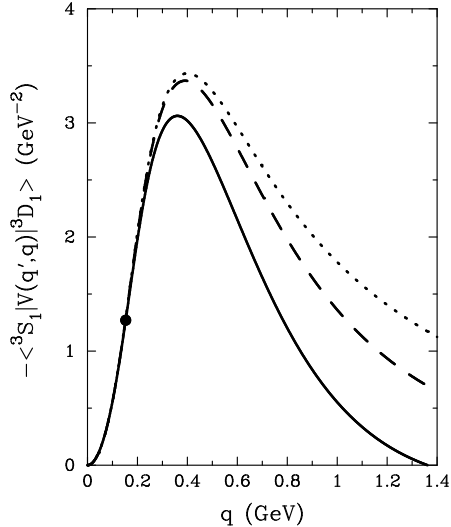,height=6.0cm}
\caption{Half off-shell $^3S_1$--$^3D_1$ amplitude
for the relativistic CD-Bonn potential (solid line), Eq.~(11).
The dashed curve is obtained when the local
approximation, Eq.~(13), is used for OPE, and the dotted curve
results when this approximation is also used for one-$\rho$
exchange. $q'=153$ MeV.}
\vspace*{10pt}
\end{figure}

Clearly, the Feynman amplitudes, Eq.~(10), are in general nonlocal
expressions; i.~e., Fourier transforming them into 
configuration space
will yield functions of $r$ and $r'$, the relative
distances between the two in- and out-going nucleons,
respectively.
The square root factors create additional nonlocality.

While 
nonlocality appears quite plausible 
for heavy vector-meson exchange (corresponding to short distances),
we have to stress here
that even the one-pion-exchange (OPE) Feynman 
amplitude is nonlocal.
This is important because the pion creates the dominant part of the 
nuclear tensor force
which plays a crucial role in nuclear structure.

Applying $\Gamma^{(\pi)}=g_\pi \gamma_5$ in Eq.~(10),
yields the Feynman amplitude for neutral pion exchange in $pp$ scattering,
\begin{eqnarray}
i{\cal A}_\pi ({\bf q'}, {\bf q})  = & -\frac{g^2_\pi}{4M^2}
\frac{(E'+M)(E+M)}
{({\bf q'}-{\bf q})^2+m_\pi^2}
\left(
 \frac{\bbox{\sigma_{1} \cdot } {\bf q'}}{E'+M}
-
 \frac{\bbox{\sigma_{1} \cdot } {\bf q}}{E+M}
\right)
\nonumber \\
 & 
\times
\left(
 \frac{\bbox{\sigma_{2} \cdot } {\bf q'}}{E'+M}
-
 \frac{\bbox{\sigma_{2} \cdot } {\bf q}}{E+M}
\right) \; .
\end{eqnarray}
This is the original and correct result for OPE.

If one now introduces the drastic approximation,
\begin{equation}
E'\approx E \approx M \; ,
\end {equation}
then one obtains the momentum space representation of the {\it local} OPE,
\begin{equation}
V_\pi^{(loc)}({\bf k})  =  -\frac{g_{\pi}^{2}}{4M^{2}}
 \frac{{\mbox {\boldmath $(\sigma_{1} \cdot $}} {\bf k)}
{\mbox{\boldmath $(\sigma_{2} \cdot $}} {\bf k)}}
 {{\bf k}^{2}+m_{\pi}^{2}}
\end{equation}
with ${\bf k} = {\bf q'} - {\bf q}$.
Notice that on-shell, i.~e., for $|{\bf q'}|=|{\bf q}|$,
$V^{(loc)}_\pi$ equals $i{\cal A}_\pi$.
Thus, the nonlocality affects the OPE potential
off-shell.

Fourier transform of Eq.~(14) yields the well-known local OPE potential
in $r$-space,
\begin{eqnarray}
V_\pi^{(loc)}({\bf r}) = &
\frac{g^2_\pi}{12\pi} 
\left(\frac{m_\pi}{2M}\right)^2
\left[ 
\left(
\frac{e^{-m_\pi r}}{r}
-\frac{4\pi}{m_\pi^2}\delta^{(3)}({\bf r})
\right)
\mbox{\boldmath $\sigma_{1} \cdot \sigma_{2}$}
\right.
\nonumber \\
& +
\left.
\left(1+\frac{3}{m_\pi r}+\frac{3}{(m_\pi r)^2}\right)
\frac{e^{-m_\pi r}}{r}
\mbox{\boldmath $S_{12}$} 
\right] \; ,
\end{eqnarray}
where $m_\pi$ denotes the pion mass.
Notice, however, that this `well-established' local OPE potential
is only an approximative representation of the correct OPE Feynman
amplitude. A QED analog is the local Coulomb potential
{\it versus} the full field-theoretic one-photon-exchange
Feynman amplitude.

\begin{table}[t]
\caption{Modern high-precision NN potentials and 
their predictions for the two- and three-nucleon systems.}
\begin{tabular}{|l|c|c|c|c|c|c|}
\hline
\hline
              & CD-Bonn~\cite{MSS96}& 
 Nijm-I~\cite{Sto94}& 
 Nijm-II~\cite{Sto94}& 
 Reid93~\cite{Sto94}& $V_{18}$~\cite{WSS95} & {\it Nature}\\
\hline
\hline
Character & nonlocal&nonloc.\ centr.\ pot.\ & local & local & local &nonlocal \\
          &          &local otherwise&       &       &       &           \\
\# of parameters& 45   & 41   & 47       & 50             & 40   & --    \\
$\chi^2$/datum  & 1.03 & 1.03 & 1.03     & 1.03           & 1.09  & --   \\
$g^2_\pi/4\pi$  & 13.6 & 13.6 & 13.6     & 13.6           & 13.6  & 14.0(5) \\
\hline
{\it Deuteron properties:} &&&&&& \\
Quadr.\ moment (fm$^2$) & 0.270 & 0.272 & 0.271 & 0.270 & 0.270  & 
   0.276(3)\tablenote{Corrected for meson-exchange 
   currents and relativity.} \\
Asymptotic D/S state & 0.0255& 0.0253 & 0.0252 & 0.0251 & 0.0250 & 0.0256(4) \\
D-state probab.\ (\%) & 4.83& 5.66&5.64 & 5.70  & 5.76     &  --    \\
\hline
{\it Triton binding (MeV):} &&&&&& \\
nonrel.\ calculation & 8.00 &7.72  & 7.62  & 7.63 & 7.62   & --     \\
relativ.\ calculation & 8.2  & --   &  -- & -- & -- & 8.48\\
\hline
\hline
\end{tabular}
\end{table}

It is now of interest to know by how much
the local approximation changes the original amplitude.
This is demonstrated in Fig.~1, where the half off-shell
$^3S_1$--$^3D_1$ potential, 
which can be produced only by tensor forces,
is shown.
The on-shell momentum $q'$ is held fixed at 153 MeV
(equivalent to 50 MeV laboratory energy),
while the off-shell momentum $q$ runs from zero
to 1400 MeV.
The on-shell point ($q=153$ MeV) is marked by a solid dot.
The solid curve is the CD-Bonn potential which contains
the full, nonlocal OPE amplitude Eq.~(12).
When the static/local approximation, Eq.~(13), is made,
the dashed curve is obtained.
When this approximation is also used for the one-$\rho$
exchange, the dotted curve results.
It is clearly seen that the static/local approximation
substantially increases the tensor force off-shell.
Clearly, we are dealing here not with negligible effects,
and the local approximation is obviously not a good one.

Even though the spirit of the new generation of potentials is
more sophistication, only the CD-Bonn potential uses
the full, original, nonlocal Feynman amplitude for OPE, Eq.~(12),
while all other potentials still apply the local approximation,
Eqs.~(14) and (15). As a consequence of this, the CD-Bonn potential
has a weaker tensor force as compared to all other potentials.
This is reflected in the predicted D-state probability of
the deuteron, $P_D$, which is due to the nuclear tensor force.
While CD-Bonn predicts $P_D=4.83$\%, the other potentials
yield $P_D= 5.7(1)$\% (cf.\ Table~2).
These differences in the strength of the tensor force lead to
considerable differences in the nuclear structure predictions
(see discussion below).

The OPE contribution to the nuclear force essentially takes care of the
long-range interaction and the tensor force.
In addition to this, all models must describe the intermediate
and short range interaction, for which very different
approaches are taken.
The CD-Bonn includes (besides the pion)
the pseudoscalar $\eta (549)$ meson, 
the vector mesons $\rho (769)$ and $\omega (783)$, and two scalar bosons
$a_0/\delta (983)$ and $\sigma (550)$, using the full, nonlocal Feynman
amplitudes, Eq.~(10), for their exchanges. 
Thus, all components of the CD-Bonn are nonlocal and the off-shell
behavior is the original one that is determined from
relativistic field theory.

The models Nijm-I and Nijm-II are based upon the Nijmegen78
potential~\cite{NRS78} (discussed in the historical section above)
which is constructed from approximate OBE amplitudes.
Whereas the Nijm-I uses the totally local approximations
for all OBE contributions, the Nijm-I keeps some nonlocal
terms in the central force component (but the Nijm-I
tensor force is totally local).
Nonlocalities in the central force have only a very
moderate impact on nuclear structure as compared to
nonlocalities in the tensor force. Thus, if for some reason one
wants to keep only some of the original nonlocalities 
in the nuclear force and not all of them,
then it would be more important to keep the tensor force nonlocalities.

The Reid93~\cite{Sto94} and Argonne $V_{18}$~\cite{WSS95} potentials do not 
use meson-exchange for 
intermediate and short range; instead, a phenomenological parametrization
is chosen.
The Argonne $V_{18}$ uses local functions
of Woods-Saxon type, 
while Reid93 applies local Yukawas of multiples
of the pion mass, similar to the original Reid potential
of 1968~\cite{Rei68}.
At very short distances, the potentials are regularized 
either by 
exponential ($V_{18}$, Nijm-I, Nijm-II) or by dipole (Reid93)
form factors (which are all local functions).

In Fig.~2, the five high-precision potentials
(in momentum space) and their phase shift predictions
are shown,
for the $^1S_0$ and $^3S_1$ states.
While the phase shift predictions are indistinguishable,
the potentials differ widely---due to the theoretical and mathematical
differences discussed. Note that NN potentials differ the most in
$S$-waves and converge with increasing $L$ (where $L$ denotes the
total orbital angular momentum of the two-nucleon system).

\begin{figure}[t]
\vspace{1.5cm}
\epsfig{file=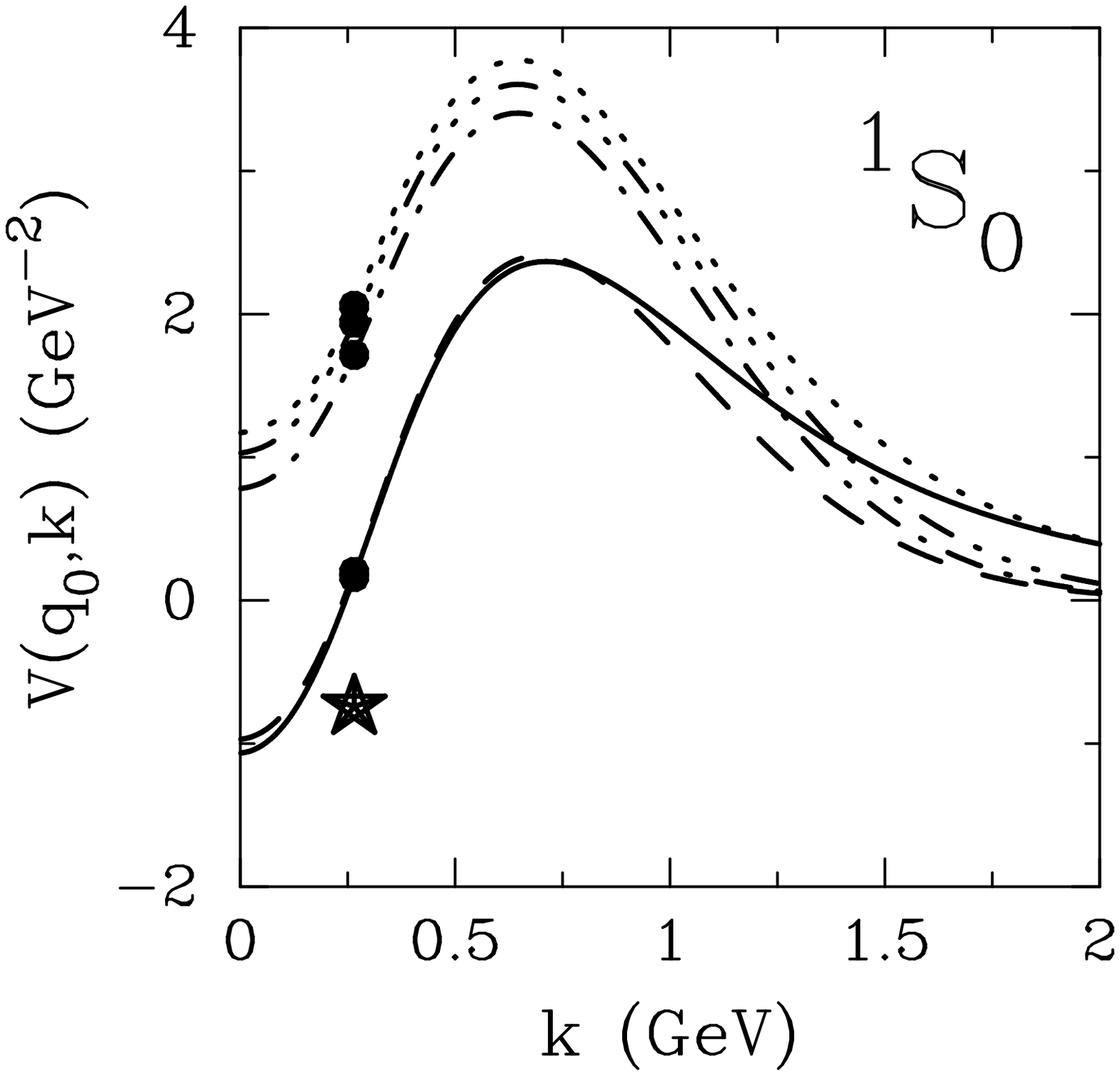,width=8cm}
\epsfig{file=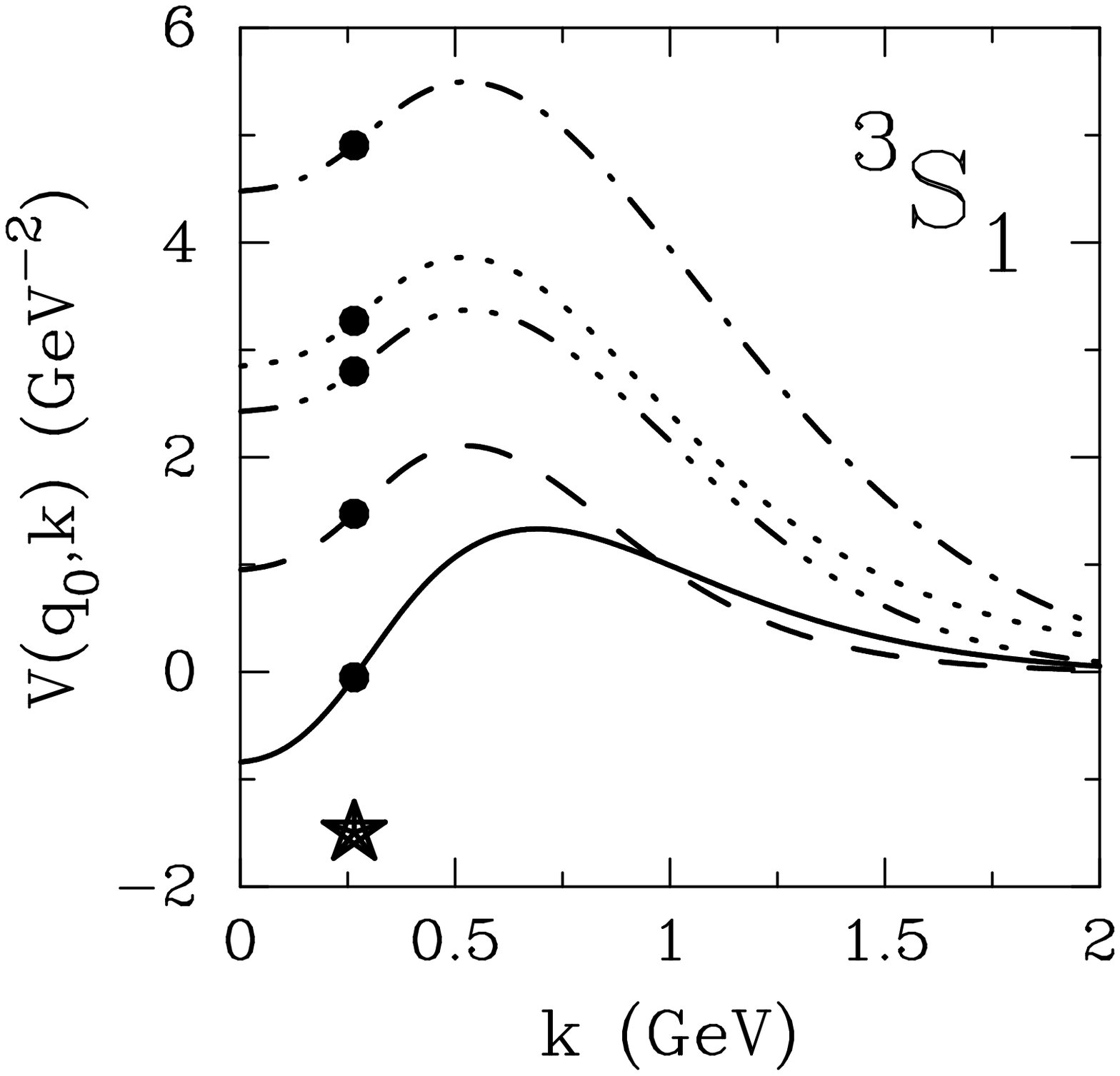,width=8cm}
\epsfig{file=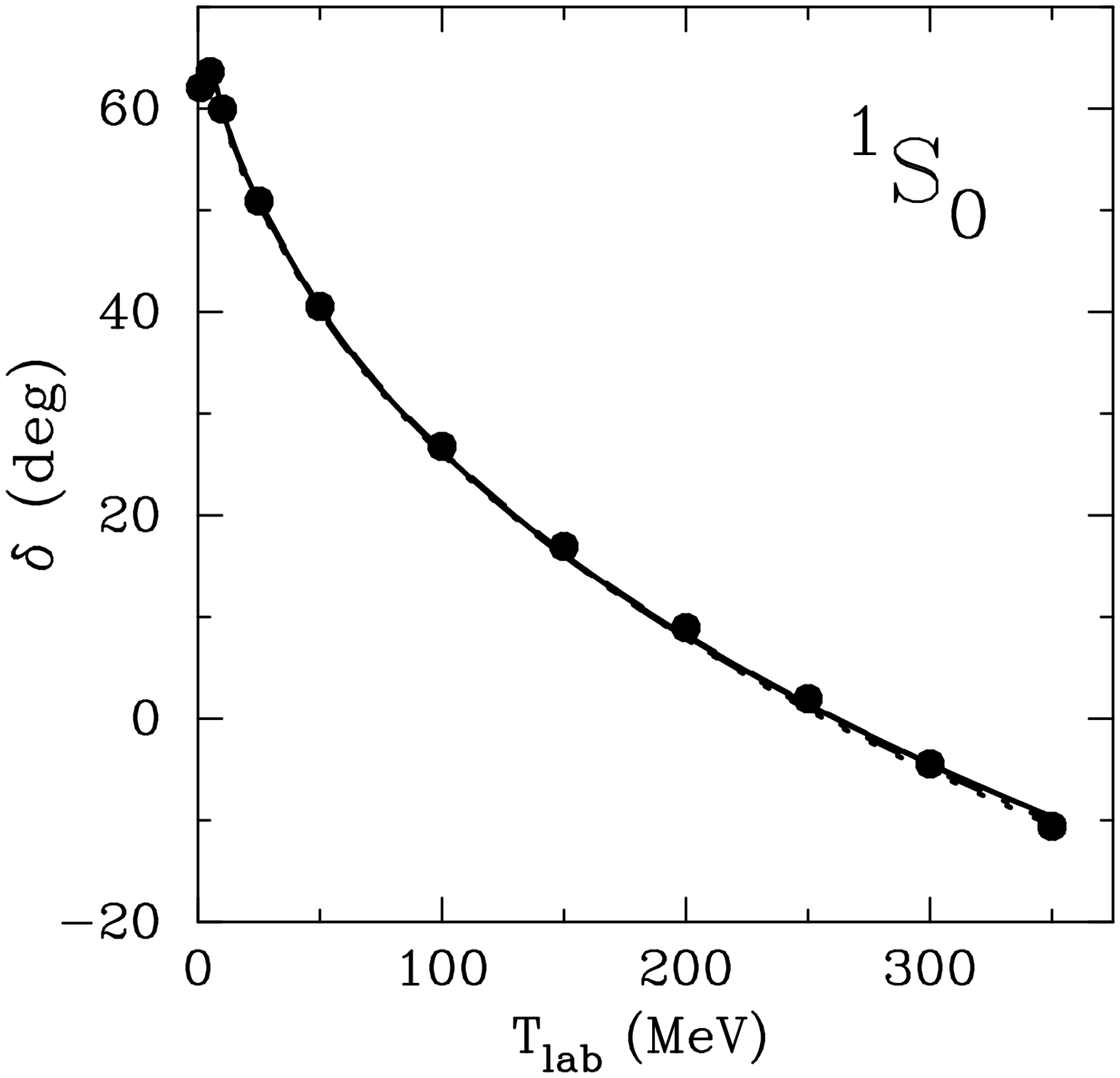,width=8cm}

\vspace*{-6.1cm}
\hspace*{8.75cm}
\epsfig{file=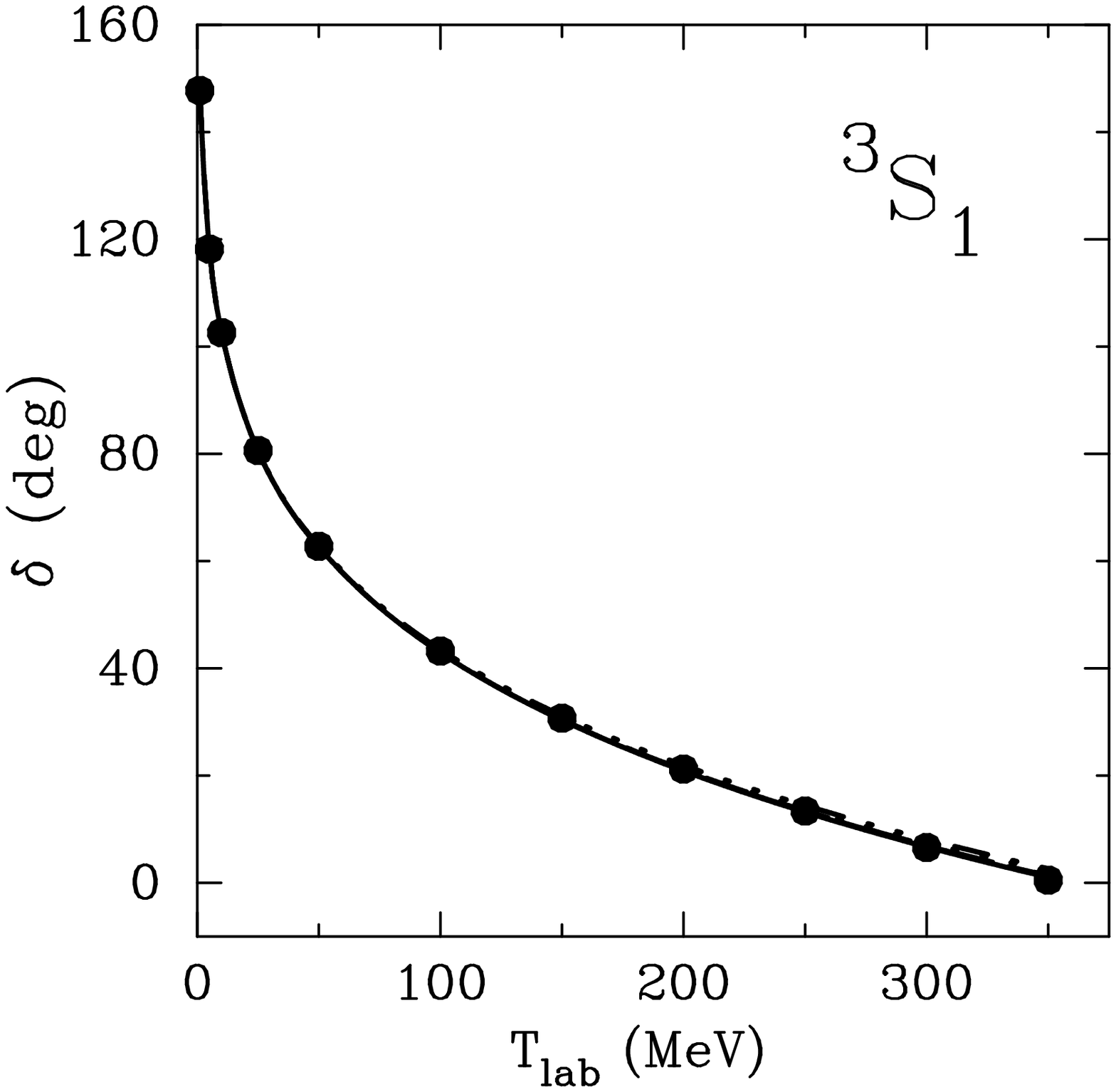,width=8cm}
\vspace*{-1cm}
\caption{{\bf Upper part:}
Matrix elements $V(q_0,k)$ of the $^1S_0$ and $^3S_1$
potentials for the 
CD-Bonn (solid line),
Nijm-I (dashed), Nijm-II (dash-dot), Argonne $V_{18}$
(dash-triple-dot) and Reid93 (dotted) potentials.
The diagonal matrix elements with $k=q_0=265$ MeV/c (equivalent
to $T_{lab}=150$ MeV) are marked by a solid dot.
The corresponding matrix element of the full scattering $R$-matrix is marked 
by the star.
{\bf Lower part:}
Predictions for the $np$ phase shifts in the $^1S_0$
and $^3S_1$ state by the five potentials.
The five curves are essentially indistinguishable. The solid dots represent
the Nijmegen multi-energy $np$ analysis [17].}
\end{figure}

Before we finish this section, a word about charge dependence
is in place. All new potentials are charge-dependent which is
essential for obtaining a good $\chi^2$. Thus, each potential
comes in three variants: $pp$, $np$, and $nn$.
The difference between $pp$ (without electromagnetic effects)
and $np$ is essentially given by the charge dependence
of OPE, Eqs.~(5) and (6), for states with $L>0$.
However, in the $^1S_0$ state, this explains only about 
50\% of the empirically known difference between 
the $pp$ and $np$ scattering lengths.
The remainder is fitted phenomenologically. 
Also the $pp$ and $nn$ singlet scattering lengths differ, which breaks
charge symmetry. The impact of the
mass difference between proton and neutron
on the kinetic energy and on the OBE amplitudes, alone,
does not explain this difference.
Therefore, the CD-Bonn and the Argonne $V_{18}$ $nn$ potentials
are phenomenologically adjusted to reproduce the $nn$ scattering length
($a_{nn}$). The potentials by the Nijmegen group (Nijm-I, Nijm-II, Reid93)
do not reproduce $a_{nn}$.

The above discussion reveals that
there is still room for improvement concerning the charge dependence
of the new high-precision potentials.
It is well-known that there are, besides OPE, other mechanisms
which create charge dependence of the nuclear force.
Recently, the charge dependence and charge asymmetry
as created by a comprehensive microscopic model have been calculated
carefully~\cite{LM98a,LM98b}. 
A refined version of the CD-Bonn potential~\cite{Mac98} 
which incorporates the results of this comprehensive study
is in preparation.
To give an idea of how important additional refinements are,
we note that the current charge-dependent, high-precision
potentials cannot explain the Nolen-Schiffer anomaly~\cite{NS69}. 
Taking the charge asymmetry derived in Ref~\cite{LM98b}
into account, however, explains the anomaly correctly~\cite{MPM98}.

\section*{NN Potentials and Nuclear Structure}

Our goal for this section is to understand how the off-shell
NN potential influences nuclear structure predictions.
For this, we need to discuss first how the NN $t$-matrix is
calculated, since it is an important quantity in the 
construction of potentials.

For a given NN potential $V$, the $t$-matrix for free-space
two-nucleon scattering is obtained from the Lippmann-Schwinger
equation, which in the center-of-mass (c.m.) frame reads
\begin{equation}
t({\bf q'}, {\bf q};E) = V({\bf q'}, {\bf q}) -
\int d^3k V({\bf q'}, {\bf k}) \frac{M}{k^2-ME-i\epsilon}
t({\bf k}, {\bf q};E)
\end{equation}
and in partial-wave decomposition
\begin{equation}
t^{JST}_{L'L}({ q'}, { q};E) = V^{JST}_{L'L}({ q'}, { q}) -
\sum_{L''} \int_0^\infty k^2 dk V^{JST}_{L'L''}({ q'}, { k})
\frac{M}{k^2-ME-i\epsilon}
t^{JST}_{L''L}({ k}, { q};E) \; ,
\end{equation}
where 
{\bf q}, {\bf k}, and ${\bf q'}$ are the relative three-momenta 
of the two interacting nucleons in the initial, intermediate, and final
state; and $q\equiv |{\bf q}|$, $k\equiv |{\bf k}|$, and $q'\equiv |{\bf q'}|$.
$E$ denotes the energy of the two interacting nucleons in the c.m. system
and is given by 
\begin{equation}
E=\frac{q_0^2}{M}
\end{equation}
with $q_0$ the magnitude of the initial relative momentum 
(c.m. on-shell momentum) which is related to
the laboratory energy by $T_{lab}=2q_0^2/M$.

Notice that the integration over the intermediate
momenta $k$ in Eqs.~(16) and (17) extends from
zero to infinity. For intermediate states with $k\neq q_0$, energy
is not conserved and the nucleons are off their energy shell (`off-shell').
The off-shell part of the potential (and of the $t$-matrix) is involved.
Thus, in the integral term in Eqs.~(16) and (17), 
the potential (and the $t$-matrix) contributes off-shell.

Note, however, that
in free-space NN scattering, 
the off-shell potential does not really play a role 
(it plays 
a significant role in the few- and
many-body problem, see below).
The reason for this is simply the procedure by which NN potentials
are constructed. The parameters of NN potentials are adjusted so
that the resulting on-shell $t$-matrix fits the empirical
NN data. 
It is important to understand this point,
for our later discussion. 
Therefore, let us consider a case for which the off-shell contributions
are particularly large, namely the on-shell $t$-matrix 
in the $^3S_1$ state:
\begin{eqnarray}
t^{110}_{00}({ q_0}, { q_0};E) & = & V^{110}_{00}({ q_0}, { q_0}) -
 \int_0^\infty k^2 dk V^{110}_{00}({ q_0}, { k}) 
\frac{M}{k^2-ME-i\epsilon}
t^{110}_{00}({ k}, { q_0};E) \nonumber \\
     &    & \mbox{}
- \int_0^\infty k^2 dk V^{110}_{02}({ q_0}, { k}) 
\frac{M}{k^2-ME-i\epsilon}
t^{110}_{20}({ k}, { q_0};E)
\end{eqnarray}
Up to second order in $V$, this is
\begin{eqnarray}
t^{110}_{00}({ q_0}, { q_0};E) & \approx & 
V^{110}_{00}({ q_0}, { q_0}) -
 \int_0^\infty k^2 dk V^{110}_{00}({ q_0}, { k}) 
\frac{M}{k^2-ME-i\epsilon}
V^{110}_{00}({ k}, { q_0}) \nonumber \\
     &    & \mbox{}
- \int_0^\infty k^2 dk V^{110}_{02}({ q_0}, { k}) 
\frac{M}{k^2-ME-i\epsilon}
V^{110}_{20}({ k}, { q_0}) \\
    & \approx &
V^{110}_{00}({ q_0}, { q_0}) 
- \int_0^\infty k^2 dk V^{110}_{02}({ q_0}, { k}) 
\frac{M}{k^2-ME-i\epsilon}
V^{110}_{20}({ k}, { q_0}) \nonumber \\
       &  &
\end{eqnarray}
where in the last equation, we have neglected the second order term in
$V^{110}_{00}$ which is, in general, smaller than the second order
in $V^{110}_{02}$.
Without partial-wave decomposition,
\begin{equation}
t({\bf q}_0, {\bf q}_0;E) \approx V_C({\bf q}_0, {\bf q}_0) -
\int d^3k V_T({\bf q}_0, {\bf k}) \frac{M}{k^2-ME-i\epsilon}
V_T({\bf k}, {\bf q}_0) \; ,
\end{equation}
where $V_C$ denotes the central force and $V_T$ the tensor force.
In words: the most important contributions are the central force
in lowest order and the tensor force in second order.

The on-shell $t$-matrix is related to the observables that are measured
in experiments.
Thus potentials which fit the same NN scattering data
produce the same on-shell $t$-matrices.
However, this does not imply that the potentials are the same.
As seen in Eqs.~(16) and (17), the $t$-matrix is the sum of two terms: the Born
term and an integral term. When this sum is the same, the individual terms
may still be quite different.

As an example, we pick up again the case of the
$^3S_1$ state, which is attractive below 300 MeV laboratory energy. 
Instead of using the $t$-matrix, it is more convenient to consider
the (real) $K$-matrix (denoted by $R$ below) which, similar to
Eq.~(19), is given by,
\begin{eqnarray}
R^{110}_{00}({ q_0}, { q_0};E) & = & V^{110}_{00}({ q_0}, { q_0}) -
 {\cal P} \int_0^\infty k^2 dk V^{110}_{00}({ q_0}, { k}) 
\frac{M}{k^2-ME}
R^{110}_{00}({ k}, { q_0};E) \nonumber \\
     &    & \mbox{}
- {\cal P} \int_0^\infty k^2 dk V^{110}_{02}({ q_0}, { k}) 
\frac{M}{k^2-ME}
R^{110}_{20}({ k}, { q_0};E) \; ,
\end{eqnarray}
where ${\cal P}$ denotes the principal value integral.

In Fig.~2 (upper part),
the on-shell $R$ matrix, $R(q_0,q_0;E)$, which corresponds to the empirical
phase shift, is denoted by a star and
the potential Born term, $V(q_0,q_0)$ is given by a solid dot,
for each potential.
The difference between  star and dot is due to the integral
term. Clearly, the size of the integral term
(in which the off-shell potential is involved) is very different
for different potentials.
Hard and strong-tensor-force potentials produce
large integral terms, while soft and weak-tensor-force
potentials produce small integral terms.
Note, however, that by construction the Born and integral terms are
balanced such that the same result is obtained for $R(q_0,q_0;E)$
(the star in Fig.~2 is the same for all potentials).

Let us now turn to the many-body problem and
consider first the three-nucleon system.
The most important point to notice here is that
a two-nucleon $t$-matrix is the input for the three-body Faddeev equations.
However, the energy parameter $E$
for this two-body $t$-matrix is different
from the one used in free-space scattering.
In the three-body Faddeev equations, one uses
\begin{equation}
E=-B_t-\frac34 \frac{q^2}{M}
\end{equation}
where $B_t$ ($\approx 8.5$ MeV) is the triton binding energy 
and $q$ is the magnitude of the momentum of the spectator nucleon.
Notice that the energy parameter $E$ is negative here, in contrast to
free scattering where $E$ is positive.
In a Faddeev calculation, $q$ runs from zero to infinity; thus,
the parametric energy for the two-body $t$-matrix ranges between
$\approx -8.5$ MeV and $-\infty$.
Inspection of Eq.~(16) reveals that this negative energy parameter
will quench the integral term of the Lippmann-Schwinger equation
due to an increase of the energy-denominator. Since
this integral term is attractive, the $t$-matrix will become less attractive
as a result of this quenching. This effect will be particularly large for the 
$t$-matrix in the $^3S_1$ state, Eq.~(19).
The consequence is that
potentials with large integral terms (i.~e., hard
potentials, large $P_D$), will experience more quenching
of the attractive integral term and, thus, lose more attraction
than potentials with a small integral term (soft potentials).

These arguments can also be stated in more intuitive form.
All NN potentials reproduce the deuteron binding energy.
In the deuteron, the tensor force couples the $^3S_1$ 
to the $^3D_1$ state. The triton is a much smaller
system than the deuteron. Therefore, in the triton, the $D$-state
is located at much higher energy than in the deutron.
This increases the energy gap between the $S$ and $D$ states and
makes an $S$-$D$ transition via the tensor force less likely
in the denser system~\cite{Ing39}. This `medium effect'
on the second order tensor contribution [cf.\ Eq.~(22)]
reduces the binding energy. The stronger the tensor force,
the larger the reduction.

This dependence of the triton binding energy predictions on the
off-shell potential, particularly
the off-shell tensor potential, is confirmed by the results
shown in the lower part of Table 2.
In a charge-dependent
34-channel momentum-space Faddeev calculation, one obtains 8.00
MeV using the CD-Bonn potential and 7.62(1) MeV applying 
any of the local potentials.
This difference of 0.38 MeV
is due to the off-shell
differences between the local and nonlocal potentials.

The unacquainted observer may be tempted to believe
that this difference of 0.38 MeV is quite small,
almost negligible. This is not true.
The discrepancy between the predictions of local potentials
(7.62 MeV) and experiment (8.48 MeV)
is 0.86 MeV.
Thus, the problem with the triton binding is that
0.86 MeV cannot be explained in the simplest way,
that is all.
Any non-trivial contribution
must, therefore, be measured against the 0.86 MeV gap between experiment
and the simplest theory.
On this scale, the nonlocality of the CD-Bonn
explains almost 50\% of the gap; i.~e., it is substantial. 

Concerning the remaining difference between theory and experiment,
two comments are appropriate. First, 
besides the relativistic, nonlocal effects that can be absorbed
into the two-body potential concept [Eq.~(11)], 
there are further relativistic corrections
that come from a relativistic treatment of the three-body system. 
This increases
the triton binding energy by 0.2--0.3 MeV~\cite{SXM92,RT92,SM98,MSS96}.
Second, notice
that the present nonlocal potentials include only the
nonlocalities that come from meson-exchange. 
However, the composite structure (quark substructure) of hadrons
should provide additional nonlocalities which may be even larger.
It is a challenging topic for future research to derive 
these additional nonlocalities, and test their impact on 
nuclear structure predictions.

\begin{table}
\caption{Predictions by some high-precision NN potentials 
for the energy per nucleon in nuclear matter (in units of MeV)
at $k_F=1.35$ fm$^{-1}$. The contributions from some important
partial waves and the totals are given.}
\begin{tabular}{crrr}
   \hspace*{3cm}      & \hspace*{2.0cm} CD-Bonn  
                      & \hspace*{2.4cm} Nijm-I 
                      & \hspace*{2.3cm}  Nijm-II \\
\hline
$^1S_0$ & --16.75 & --16.73 & --16.11 \\
$^3S_1$ & --19.00 & --17.55 & --17.05 \\
$^3P_0$ & --3.09 & --3.11 & --3.06 \\
$^3P_1$ & 9.84 & 9.73 & 9.72 \\
$^3P_2$ & --7.03 & --7.00 & --7.01 \\
Total potential energy   & --36.35 & --35.07 & --33.70 \\
Kinetic energy   & 22.67 & 22.67 & 22.67 \\
Total energy   & --13.68 & --12.40 & --11.02 \\
\end{tabular}
\end{table}

For heavier systems, the Brueckner $G$-matrix
is the basic quantity for all
calculations of nuclear ground and excited states.
The $G$-matrix is the solution of the
Bethe-Goldstone equation,
\begin{equation}
G({\bf q'}, {\bf q}) = V({\bf q'}, {\bf q}) -
\int d^3k V({\bf q'}, {\bf k}) \frac{M^\star Q}{k^2-q^2}
G({\bf k}, {\bf q}) \; ,
\end{equation}
which differs from the Lippmann-Schwinger equation, Eq.~(16),
in two points:
the Pauli projector $Q$ and the energy denominator.
(For simplicity, I assume in Eq.~(25) 
the so-called continuous choice
for the single particle energies in nuclear matter, i.~e., the energy
of a nucleon is represented by $\epsilon(p) = p^2/(2M^\star) - U_0$
for $p\leq k_F$ as well as $p> k_F$, where $M^\star$ is the effective
mass and $U_0$ a positive constant.)
The Pauli projector prevents scattering into occupied states and, thus,
cuts out the low momenta in the $k$ integration.
The change introduced by the Pauli projector is known as the Pauli effect.
The energy denominator gives rise to the so-called dispersive effect.
When using the continuous choice, the dispersive effect is given simply
by the replacement of $M$ by $M^\star$ ($\approx \frac23 M$ at nuclear
matter density) in the numerator of the integral
term, which leads to a reduction of this attractive term
by a factor $M^\star/M$.
Both effects are in the same direction, namely they quench the
integral term. Since the integral term is negative, these effects
are repulsive.

In Table 3, nuclear matter results are given for three 
representative high-precision potentials using the Brueckner
pair approximation and the conventional choice for the
single particle potential. For the reasons discussed,
the largest difference between the predictions occurs in
the $^3S_1$ state where the tensor force is involved.
In the $^1S_0$ state, only the central force can contribute
which is nonlocal for CD-Bonn and Nijm-I.
This explains why the binding energy predicted for this state
is the same for these two potentials, and is larger than the value predicted by
the local potential Nijm-II.

Note that there are further contributions in nuclear matter
beyond the Brueckner pair approximation. 
Three- and four-body clusters contribute
about 4 MeV attraction~\cite{Son98} and the medium effect
on the Dirac spinors representing the nucleons in nuclear
matter creates about 2 MeV repulsion~\cite{Mac89}
resulting in a net correction of about --2 MeV. 
This brings the prediction of the nonlocal potential into
good agreement with the empirical nuclear matter
value of $-16\pm 1$ MeV.

The trend of the nonlocal Bonn potential to increase binding
energies has also a very favorable impact on predictions for
spectra of open-shell nuclei~\cite{Jia92,And96}.

\section*{Three-Nucleon Forces}

\begin{figure}[t]
\vspace*{-3.0cm}
\hspace*{-1.0cm}
\epsfig{file=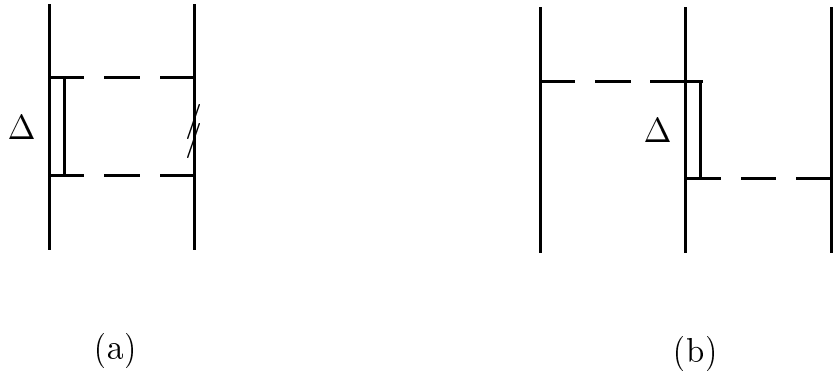,width=20.0cm}
\vspace*{-16.5cm}
\caption{Two- and three-body forces created by the $\Delta$ isobar.
Solid lines represent nucleons, double lines $\Delta$ isobars,
and dashed lines $\pi$ and $\rho$ exchange.
{\bf Part (a)} is a contribution to the two-nucleon force.
The double slash on the intermediate nucleon line is to indicate
the medium modifications (Pauli blocking and dispersion effects)
that occur when this diagram is inserted into a nuclear many-body
environment.
{\bf Part (b)} is a three-nucleon force.}
\end{figure}

Strictly speaking, most many-body forces are an artifact of theory.
They are created by `freezing out' degrees of freedom contained
in the full Hamiltonian of the problem.
This fact allows us to derive a guideline for dealing 
with the many-body-force issue in a consistent way:
when you introduce a new degree of freedom,
do not freeze it out; instead
take it into account in the two- 
{\it and} many-body problem, consistently. 
Field-theoretic models for the $2\pi$-exchange contribution to
the NN interaction require the $\Delta(1232)$ isobar, which
also creates a three-nucleon force (3NF) in the many-body system.
Consequently, when introducing the isobar degree of freedom, it should
be included in the two- and three-body force simultaneously.
When these forces are applied in the many-body problem, then
one is faced with two effects: a (repulsive) medium effect on the two-body
force [Fig.~3(a)] and an attractive 3NF contribution [Fig~3(b)].
The repulsive medium effect has been known for more than 20 years~\cite{HM77}.
{\it Consistency requires that either both effects
are taken into account or none.}
Consistent calculations of this kind have first been conducted
by the Hannover group~\cite{HSS83} for the triton. Recently, these
calculations have been improved and extended by Picklesimer
and coworkers~\cite{Pic92} using the Argonne $V_{28}$ 
$\Delta$-model~\cite{WSA84}.
They find an attractive contribution to the triton energy
of --0.66 MeV from the 3NF diagrams created by the $\Delta$,
and a repulsive contribution of +1.08 MeV from the dispersive effect
on the two-nucleon force involving $\Delta$ isobars.
The total result is {\it 0.42 MeV repulsion}~\cite{Pic92}.

A contribution that is missing in the $\Delta$ 3NF model is the S-wave part
of the off-shell $\pi N$ amplitude. Estimates for this contribution
vary between --0.2 and --0.6 MeV~\cite{Fri88}.
Adding this to the Argonne $V_{28}$ result 
[i.~e., $0.42 - (0.4\pm 0.2)=0.0\pm 0.2$ MeV]
yields a vanishing
result for the total contribution from three-body forces.

In the more traditional calculations, 
it was customary to employ the so-called 2$\pi$-exchange 
three-nucleon force. This approach does not  
take the dispersive 
effects into account and overbinds the triton,
at least if commonly accepted values for the
cut-off parameter of the strong pion-nucleon vertex are used.
Moreover, this type of 3N force fails in 3N scattering~\cite{WHG94} 
casting additional doubt on its reality.

Besides the $\Delta$-isobar, there are other mechnisms that can
give rise to three-nucleon forces. In recent years, it has become
fashionable to consider chiral Lagrangians for nuclear
interactions. Such Langrangians may create diagrams which
represent effective three-nucleon potentials. 
However, Weinberg has shown that all these diagrams cancel~\cite{Wei90}.

In conclusion, any {\it consistent} three-body force calculation---conducted
to date---has yielded vanishing results.
This implies that the two-body force
should provide essentially all binding energies observed
in nuclei. The further implication then is that
nonlocal/weak-tensor force NN interactions are to be favored
over hard/local potentials.
It is probably not just an accident that these
soft/nonlocal potentials are also more
faithful to the underlying theory of nuclear forces.

\section*{Summary and Outlook}

Several high-quality/high-precision NN potentials are now available
which fit the low-energy NN data with identical perfection.
These potentials differ, however, in their off-shell
behavior. Thus, the stage is set for a reliable investigation
of off-shell effects in microscopic nuclear structure calculations.
Such calculations may finally teach us something about the
off-shell nature of the nuclear force.

This is {\it the} time to do {\it microscopic} nuclear structure calculations!

\section*{Acknowledgement}
This work was supported in part by the U.S.\
National Science Foundation under Grant-No.~PHY-9603097.


\begin{references}
\bibitem{foot1} A more detailed 
discussion of this issue and figures illustrating the facts 
can be found in Ref.~\cite{ML94}.
\end{references}
\end{document}